# Crystal structure instability of FeSe grains: Formation of non-superconducting phase at the grain surface

Hiroki Izawa, Yuji Tanaka, Yoshikazu Mizuguchi*, and Osuke Miura

Department of Electrical and Electronic Engineering, Tokyo Metropolitan University, Hachioji, Tokyo 192-0397, Japan

*E-mail : mizugu@tmu.ac.jp

**Abstract**

We investigated the magnetization and crystal structure of FeSe polycrystalline samples with various grain sizes. For the samples with large grains, a large magnetic critical current density $J_c$ was observed. For the samples with small grains, superconductivity signals were not observed; instead, magnetization hysteresis, which is not a result of superconductivity, was observed. In the X-ray diffraction pattern for the samples with small grains, broad additional peaks were observed, corresponding to the formation of the non-superconducting (monoclinic) Fe-Se phase at the FeSe grain surface. The crystal structure instability at the grain surface would be the reason for the low superconducting properties of the Fe-chalcogenide polycrystalline wires investigated thus far.

## 1. Introduction

The discovery of superconductivity in Fe-based superconductors has been drawing considerable attentions in the field of condensed matter physics owing to their unconventional superconductivity mechanisms and high transition temperature ($T_c$) as high as 58 K [1-4]. Furthermore, in the field of superconductivity application, Fe-based superconductors have been considered as potential materials for superconducting application owing to their high performance characteristics, such as high $T_c$, high upper critical field ($H_{c2}$), and small anisotropy of superconductivity [4]. Owing to their high superconductivity performance at high magnetic fields, one of the potential applications of Fe-based superconductors is a superconducting wire or tape for magnets. Thus far, several types of Fe-based superconducting wires have been fabricated using $A_{1-x}K_xFe_2As_2$ (A = Ba, and Sr) [5-8], $SmFeAsO_{1-x}F_x$ [9-11], and Fe chalcogenides (FeSe-based [12-19] or FeTe-based [20-24] superconductors). The FeAs-based superconducting wires (particularly, $Ba_{1-x}K_xFe_2As_2$ and $Sr_{1-x}K_xFe_2As_2$ wires) fabricated by the powder-in-tube (PIT) method showed a high critical current density ($J_c$). Recently, a high transport $J_c$ exceeding 0.1 MA/cm$^2$ under 10 T at 4.2 K has been achieved in the $Sr_{1-x}K_xFe_2As_2$ tape [8]. However, such a high transport $J_c$ has not been achieved in the Fe-chalcogenide PIT wires (or tapes) with polycrystalline cores. The highest recorded transport $J_c$ among the Fe-chalcogenide wires is about 1 kA/cm$^2$ under the self-field at 4.2 K in the FeSe wire fabricated by the in situ PIT method [14].

On the other hand, Fe chalcogenides exhibited some advantages in superconductivity applications. First, both the anisotropy of superconductivity and the toxicity of the Fe chalcogenides were lower than those of the FeAs-based superconductors. Second, high $J_c$ values were achieved in high-quality single crystals or thin films [4,25,26]. For example, transport $J_c > 10^6$ A/cm$^2$ (self-field) and $> 10^5$ A/cm$^2$ under 30 T at 4.2 K were observed in high-quality Fe(Se, Te)-coated conductors [26]. Third, the superconducting properties of the Fe-chalcogenide superconductors were markedly enhanced by tuning their crystal structure. For example, the $T_c$ of FeSe exceeded 30 K when high pressures were applied to FeSe [27-29], or K ions were intercalated into the FeSe layers ($K_xFe_{2-y}Se_2$) [13]. Furthermore, recent studies revealed that the $T_c$ of FeSe monolayer thin films can be 50-100 K [30-32]. On the basis of these high potential abilities, the development of the fabrication process for high-performance Fe-chalcogenide superconducting wires or tapes is highly required.

In the fabrications of many types of FeSe [and Fe(Se,Te)] wires and tapes thus far, we have noted that the superconducting properties were largely affected by the grinding of the precursor powders and/or the drawing of the wires. The superconducting transition, which was clearly observed in the precursor samples, sometimes disappeared (or degraded) after the drawing process. To clarify the origin of this phenomenon and to propose the optimal method for the fabrication of Fe-chalcogenide wires or tapes, we systematically investigated the superconducting (magnetic) properties, grain size, and crystal structure of FeSe powders with different grain sizes. We show that decreasing the grain size results in the formation of the monoclinic FeSe phase, which is not superconducting but magnetically ordered, at the grain surface. As a conclusion, we propose that the formation of monoclinic phases would be the reason for the low $J_c$ in the polycrystalline Fe-chalcogenide wires and tapes, and that a high $J_c$ can be obtained in single crystals or thin films.

**2. Experimental methods**

Polycrystalline samples of FeSe were prepared by the melting method using Fe powders (99.9%) and Se chips (99.999%). The starting materials with a nominal composition of Fe:Se = 1:1 were double-sealed into evacuated quartz tubes and heated at 1100 °C for 10 h. After furnace cooling, the product was annealed at 400 °C for 200 h, followed by quenching to the room temperature: in this article, we call this sample *As-prepared* sample. The FeSe powders with various grain sizes were prepared by grinding the obtained sample (As-prepared sample) using an agate mortar at various grinding times of 1, 15, and 30 min: these powder samples are called *1-, 15-, and 30-min powders*, respectively. As a reference, the monoclinic Fe-Se phase ($Fe_3Se_4$) was also synthesized by the solid-state reaction method using the Fe powders and Se chips. The temperature dependence of magnetization was measured using a superconducting quantum interference device (SQUID) magnetometer down to 2 K after zero-field cooling with an applied field of 10 Oe. The magnetization loop (the magnetic field dependence of magnetization) was measured up to 7 T at 4.2, 10, and 300 K. The powder X-ray diffraction (XRD) was performed using the $2\theta$-$\theta$ method within a range of $2\theta$ = 10–80 deg with Cu-K$\alpha$ radiation. The XRD pattern for the 1-min powder was refined using the Rietveld method with the RIETAN-FP program [33] to investigate the secondary phase and lattice constants. The surface and grain size of the samples were investigated by scanning electron microscopy (SEM).

**3. Results and discussion**

Figure 1 displays the SEM images of all the samples at two different magnifications (×1000 and ×10000). For the As-prepared sample, large grains with step structures, which are typically observed in Fe-chalcogenide single crystals, are observed. For the 1-min powder, large grains with a typical grain size of 10-20 μm remain. For the 15-min powder, such large grains are not observed, although small grains with a typical grain size of 0.2-1 μm are observed. For the 30-min powder, the size of most grains is less than 0.1 μm. These SEM images confirm that we achieved systematic preparations of FeSe powders with different grain sizes.

Figure 2(a) shows the temperature dependence of magnetization for the As-prepared FeSe sample and the FeSe powders prepared by grinding for 1, 15, and 30 min. The superconducting transition ($T_c$ ~ 9 K) with a large shielding volume fraction is observed for the As-prepared sample and 1-min powder, indicating that the samples show bulk superconductivity. In contrast, the 15- and 30-min powders do not show large diamagnetic signals, suggesting that the bulk characteristics of superconductivity are almost lost in these powder samples.

Figures 2(b) and 2(c) show the magnetization loops of the As-prepared sample and 1-min powder, respectively. For the As-prepared sample, the magnetization hysteresis is observed at $T = 4.2$ K only. Since the hysteresis disappears above 10 K ($> T_c = 9$ K), this hysteresis is considered to result from the superconductivity. For the 1-min powder, the magnetization loops similar to those of the As-prepared sample are observed while a small hysteresis curve is observed at 10 K as well. This suggests that the major phase is superconducting, but a small magnetic phase is generated in the 1-min powder. For these two samples, the magnetic field dependence of $M$ (4.2 K) – $M$ (10 K), the difference in magnetization between $T = 4.2$ and 10 K, shows the typical magnetization hysteresis of the FeSe superconductor (see the insets) [12].

Figures 2(d) and 2(e) show the magnetization loops of the 15- and 30-min powders, respectively. For these samples, the magnetization loops exhibit hysteresis curves at both 4.2 and 10 K. These hysteresis loops are not observed at 300 K. Since the $T_c$ of the present FeSe phase is 9 K, these hysteresis curves should not be attributed to the superconductivity. In addition, the disappearance of the superconducting signals in the temperature dependence of magnetization in Fig. 2(a) suggests that the superconducting $J_c$ of these samples is almost zero. On the basis of the results above, we considered that the superconducting phase (tetragonal phase) was destroyed by grinding, and a different phase, which is magnetic, was formed. In addition, we noted that the magnitude of the magnetization loop at 300 K (for example, see the values at 7 T) systematically increases with increasing grinding time as shown in the online supplementary data (Fig. S1). This suggests that the amount of the magnetic phase is increased by increasing the grinding time.

To investigate the changes in crystal structure, we performed powder XRD experiments. Figure 3 shows the XRD patterns of the 1-, 15-, and 30-min FeSe powders. For the 1-min powder, the XRD pattern corresponds to that typically observed in tetragonal FeSe [space group: *P*4/*nmm* (#129)]. The sample is almost-single-phase tetragonal FeSe with the lattice constants $a = 3.7747(3)$ Å and $c = 5.5297(6)$ Å, while small impurity peaks are detected in the hexagonal FeSe phase [space group: *P*6$_3$/*mmc* (#194)]. With increasing grinding time, the XRD peaks become broader, which is due to the decrease in the grain size. However, additional broad peaks indicated with asterisks appear in the 30-min pattern. These peaks cannot be explained with the tetragonal *P*4/*nmm* model or the decrease in grain size. On the basis of the observed additional (asterisk) peaks and the results of the magnetization measurements, we considered that the magnetization hysteresis curves observed above $T_c$ in the 15- and 30-min powders result from the new phase detected in the XRD patterns.

The broad XRD peaks of the new phase can be explained by the monoclinic Fe-Se phase with the space group of *C*2/*m* (#12). As shown in the online supplementary data (Fig. S2), the XRD pattern of the monoclinic Fe-Se phase (Fe$_3$Se$_4$) shows multiple-peak profiles at around $2\theta = 33$ and 43 deg. These peak positions are consistent with the broad peaks of the new phase, but they don't correspond to those of the tetragonal or hexagonal structure. Therefore, we measured the magnetic properties of the monoclinic $Fe_3Se_4$ polycrystalline sample to confirm that the magnetic hysteresis curves observed above $T_c$ in the 15- and 30-min powders can be attributed to the monoclinic phase. Figure 4 shows the temperature dependence of magnetization for monoclinic $Fe_3Se_4$. An anomaly is observed at around 250 K. The insets shows the magnetization loops for monoclinic $Fe_3Se_4$ at 4.2, 10, and 300 K. As we expected, magnetization hysteresis curves, which are similar to those observed in the 15- and 30-min powders, are observed. The hysteresis almost disappears at 300 K. The hysteresis can be observed below the anomaly temperature: the magnetization loops at different temperatures are shown in the online supplementary data (Fig. S3).

      We revealed the formation of the monoclinic Fe-Se phase from the magnetization and XRD measurements. Next, we determined the possible explanation of the formation of the monoclinic phase by decreasing the grain size of FeSe. In the Fe-Se binary phase diagram, there are many possible compositions and crystal structures [34]. In addition, the high sensitivity of the crystal structure to the stoichiometry was reported in the vicinity of FeSe (Fe:Se = 1:1) [35]. Having considered these facts on the structure instability, we assume that the tetragonal (superconducting) FeSe structure partially transforms to the monoclinic Fe-Se structure through the 15- and 30-min grinding processes. In addition, the broad peaks of the monoclinic phase should be suggestive of the very low crystallinity of the newly formed monoclinic Fe-Se phase. Therefore, we conclude that the monoclinic Fe-Se phase forms at the FeSe grain surface. This scenario is consistent with the fact that the signal of the monoclinic phase (in the magnetization and XRD measurements) is enhanced in the 15- and 30-min powders because the surface area is expected to increase with decreasing grain size. In that case, the composition of the monoclinic Fe-Se phase formed at the tetragonal FeSe grain surface should be important, but the analysis of the composition of the phase is very difficult. One possibility is that the composition remains to be Fe:Se = 1:1 in the monoclinic structure as well. Another possibility is that the composition deviates from 1:1 to a ratio close to $Fe_3Se_4$. The composition analysis of the nanoscale surface region should be important to further understand the mechanisms of the formation of the monoclinic structure.

      The magnetic properties of the nanostructured $Fe_3Se_4$ were reported in Refs. 35 and 36. The $Fe_3Se_4$ phase exhibits ferromagnetic transition at 315 K and exhibits a large coercivity (exceeding 40 kOe at 10 K). The emergence of ferromagnetism in $Fe_3Se_4$ below 300 K is consistent with our magnetization field curves of the 15- and 30-min samples. The amplitudes of the hysteresis loops seem to vary between our results and the results indicated in Refs. 35 and 36; for example, the 15- and 30-min samples do not exhibit magnetization hysteresis at 300 K. We consider that the difference is caused by the difference in the composition of the monoclinic phase. As mentioned above, the monoclinic phase is considered to be formed by the local structure distortion from the FeSe tetragonal phase.

Therefore, the composition could deviate from $Fe_3Se_4$ but close to FeSe. The deviation of the composition from $Fe_3Se_4$ could be the origin of the different magnetic properties. Furthermore, the shape of the monoclinic phase should affect the magnetic properties. Indeed, as shown in Figs. 2 and 4, the shapes of the magnetization loops clearly vary in these samples (nanoscale for Fig. 2 but bulk $Fe_3Se_4$ for Fig. 4). This finding suggests that the magnetic properties could be affected by the grain shape and size.

Finally, on the basis of the conclusion regarding the instability of the tetragonal phase at the FeSe grain surface, we mention the superconducting wire fabrication strategy suitable for achieving a high $J_c$ with the FeSe superconductor. As we mentioned in the introduction part, the Fe-chalcogenide wires fabricated by several PIT methods have never shown a large $J_c$. The low superconducting performance in the Fe-chalcogenide polycrystalline wires may result from the formation of the non-superconducting phase at the grain surface. Actually, a relatively large transport $J_c$ was observed in the FeSe wires fabricated at high temperatures, at which the wire core was expected to melt [14-16]. In these wires, the grain size of the FeSe core should be large as observed in the As-prepared sample of this study. However, the superconducting properties of polycrystalline FeSe are essentially low possibly owing to the structure instability between the tetragonal, hexagonal, and monoclinic structures. In addition, the high-$T_c$ superconductivity in FeSe is only induced at high pressures of 4-6 GPa, which can be understood by the optimization of the local crystal structure of the tetragonal FeSe [27-29,35,38]. These findings suggest that the crystal structure of FeSe at ambient pressure is not preferable for the emergence of Fe-based superconductivity. Hence, the transport $J_c$ achieved in the polycrystalline FeSe wires was clearly lower than that of the FeAs-base wires. As compared with those of FeSe, the superconducting properties of Fe(Se,Te) are essentially high. However, the melting process cannot be used for the Fe(Se, Te) wire because the superconducting properties of the Fe(Se, Te) compound are highly sensitive to the inclusion of the excess Fe ions at the interlayer site [23,24]. (The use of the Fe sheath results in the introduction of Fe from the sheath to the core, and the use of the other metal sheaths results in the reaction of metal tellurides, which finally results in the increase in the amount of excess Fe in the Fe(Se,Te) core [37].) Having considered the low superconducting properties (particularly in the polycrystalline PIT wires) and high superconducting properties in single crystals [25] or coated conductors [26], we propose that the best way to achieve practical performance in the Fe-chalcogenide superconducting wires (or tapes) is growing well-connected large grains as coated conductors.

4. **Conclusions**

We investigated the magnetic properties (including superconducting properties) and crystal structure of FeSe polycrystalline samples with various grain sizes. The grain size of the samples was systematically tuned by changing the grinding time (1, 15, and 30 min). For the samples with large grains (As-prepared and 1-min powder), the magnetization hysteresis corresponding to the emergence of the superconducting current was observed. For the samples with small grains with a submicron order, superconductivity signals disappeared; instead, magnetization hysteresis, which does not result from the superconducting current, was observed. The magnetization hysteresis above the superconducting $T_c$ and the appearance of new broad peaks in the XRD pattern suggest the formation of a non-superconducting (magnetic) phase with a monoclinic structure. On the basis of the evolution of the magnetization hysteresis and broad XRD peaks with decreasing grain size, we concluded that the non-superconducting (monoclinic) Fe-Se phase formed at the FeSe grain surface. Our conclusion suggests that *polycrystalline* superconducting wires (namely, PIT wires and tapes) with Fe-chalcogenide superconductors cannot achieve a high $J_c$, and that the fabrication process for achieving a high $J_c$ involves the use of single-crystalline tapes such as coated conductors.

**Acknowledgement**

This work was partly supported by Grants-in-Aid for JSPS Fellows (15J04746) and Scientific Research on Innovative Areas (15H05886).

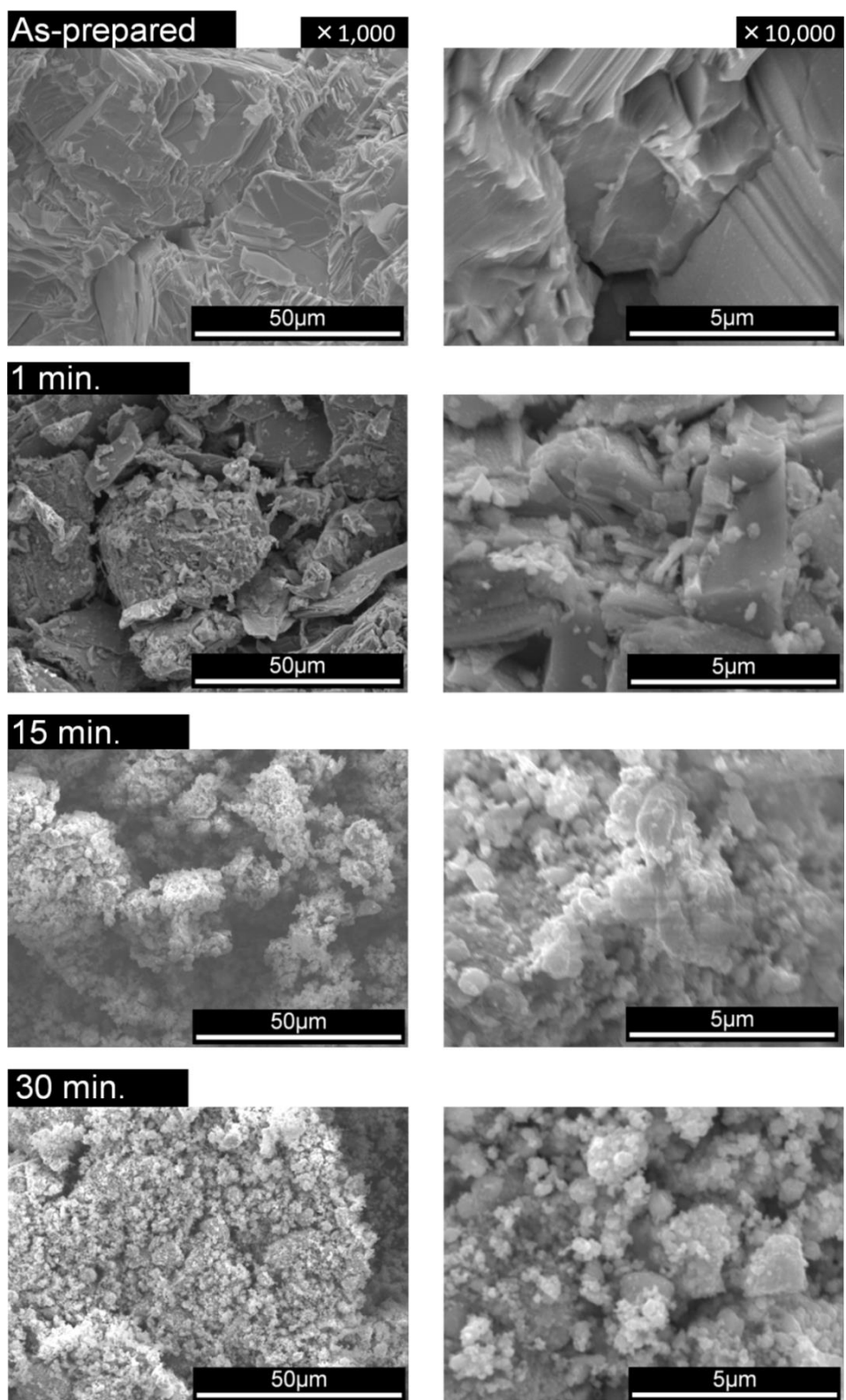

Fig. 1. SEM images of the As-prepared FeSe sample and the FeSe powders prepared by grinding for 1, 15, and 30 min.

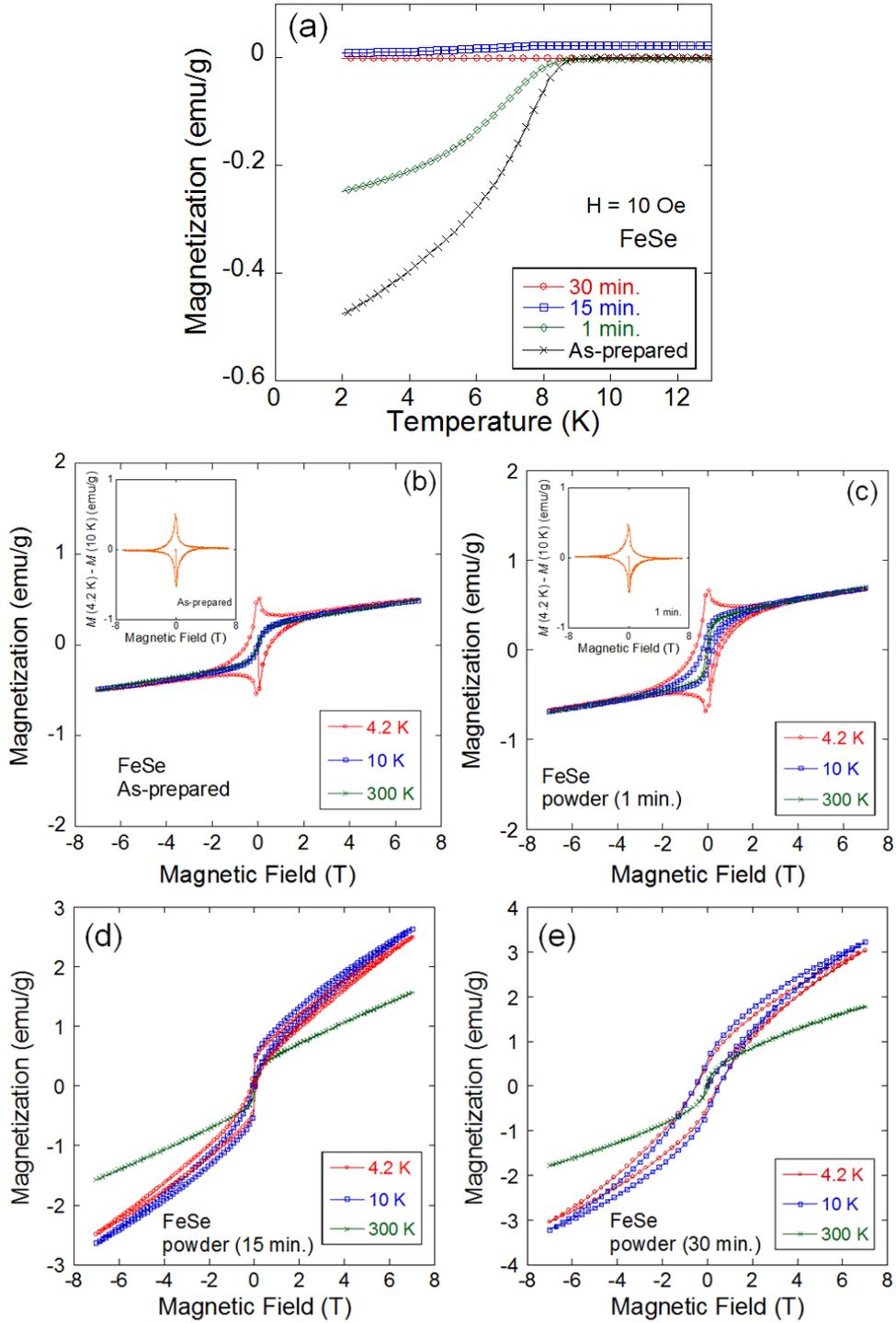

Fig. 2. (a) Temperature dependence of magnetization for As-prepared FeSe and the FeSe powders ground under various conditions (ground for 1, 15, and 30 min). (b-e) Magnetization loops for the As-prepared FeSe sample and 1-, 15-, and 30-min powders. The insets in Fig. 2(b) and 2(c) show the magnetic field dependence of the difference in magnetization between 4.2 K (below $T_c$) and 10 K (above $T_c$), $M$ (4.2 K) – $M$ (10 K).

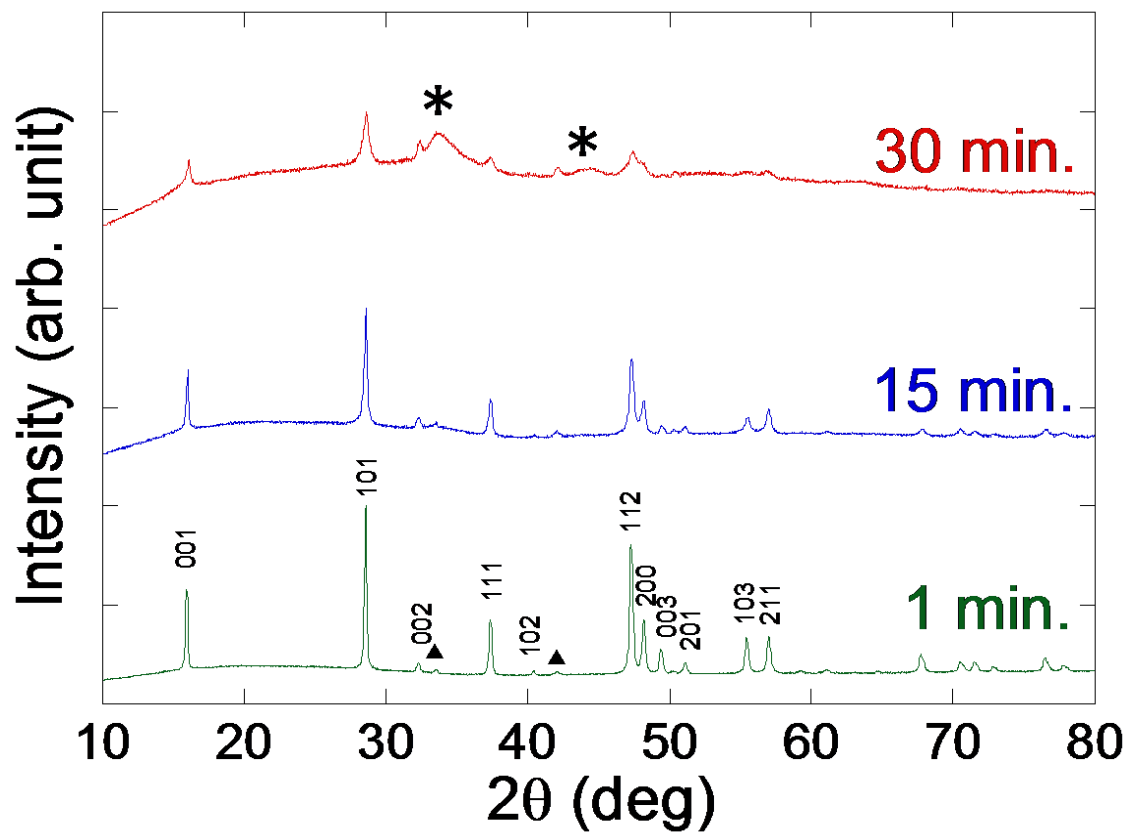

Fig. 3. XRD patterns of the 1-, 15-, and 30-min FeSe powders. The numbers in the bottom XRD pattern (1 min) indicate Miller indices for tetragonal (*P4/nmm*) FeSe. The filled triangles indicate the XRD peaks of the hexagonal FeSe (impurity phase). The asterisks indicate the expected peak positons of the monoclinic Fe-Se ($Fe_3Se_4$) phase. (The XRD pattern of the monoclinic $Fe_3Se_4$ is shown in supplementary data.)

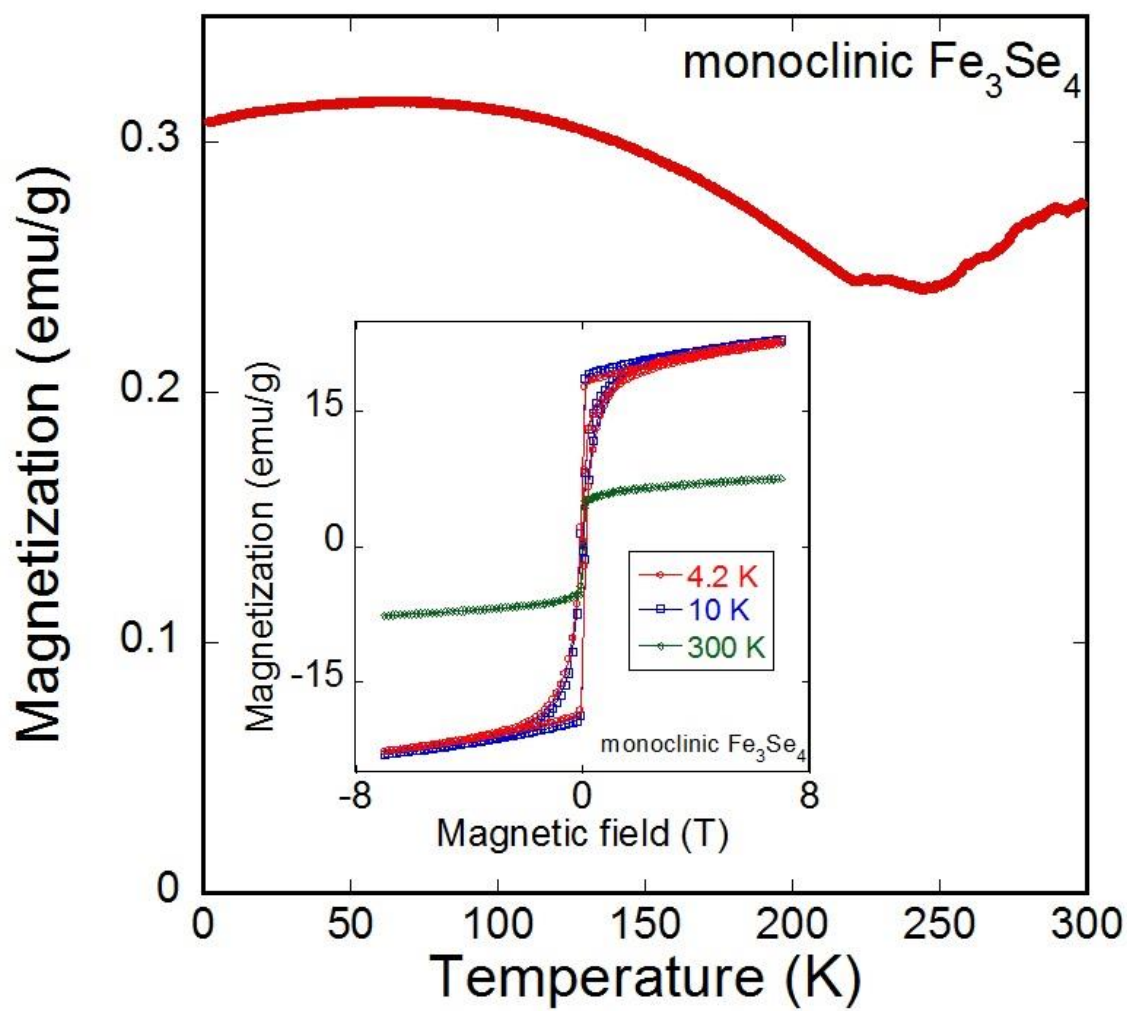

Fig. 4. Temperature dependence of magnetization for monoclinic Fe-Se ($Fe_3Se_4$). The inset shows the magnetization loop for monoclinic Fe-Se ($Fe_3Se_4$) at 4.2, 10, and 300 K. (The magnetization loops measured at other temperatures are shown in the supplementary data.)

# Crystal structure instability of FeSe grains: formation of non-superconducting phase at the grain surface

Hiroki Izawa, Yuji Tanaka, Yoshikazu Mizuguchi*, and Osuke Miura

*Department of Electrical and Electronic Engineering, Tokyo Metropolitan University, 1-1 Minami-Osawa, Hachioji 192-0397, Japan*

Corresponding author: Y. Mizuguchi (mizugu@tmu.ac.jp)

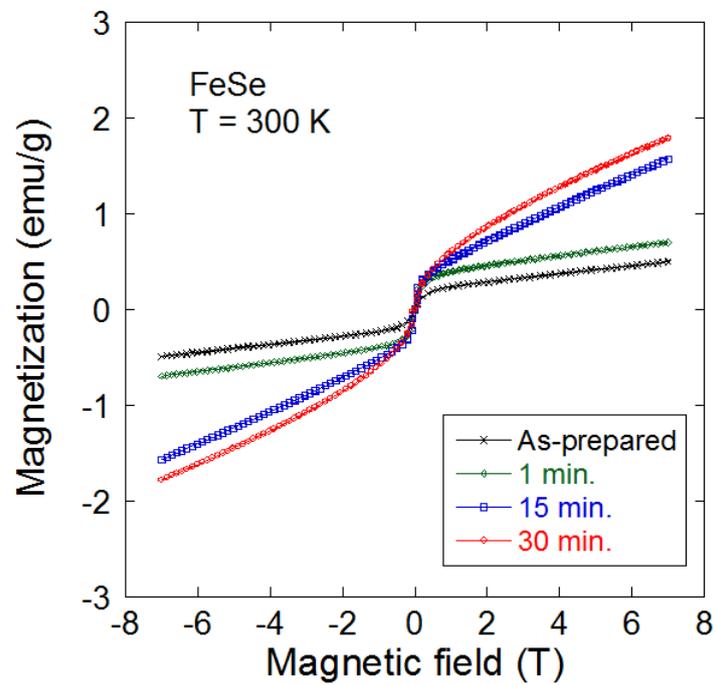

Fig. S1. Magnetization loops of FeSe (As-prepared, 1-min, 15-min, and 30-min samples) at 300 K.

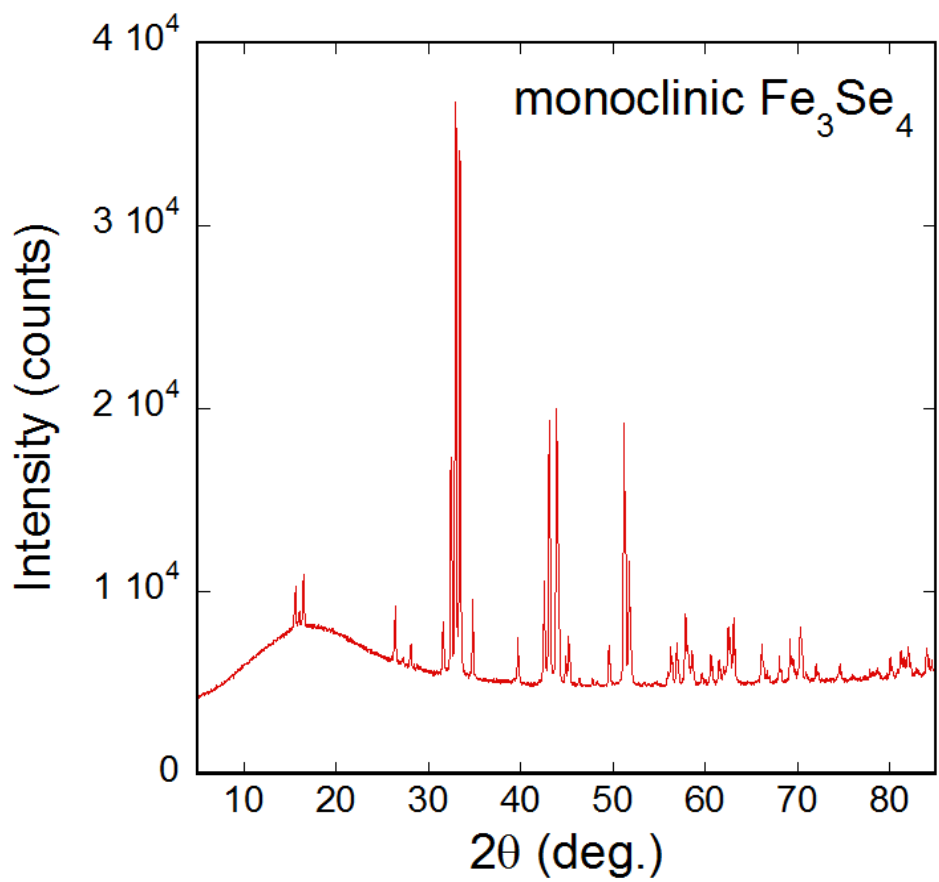

Fig. S2. XRD pattern for monoclinic $Fe_3Se_4$.

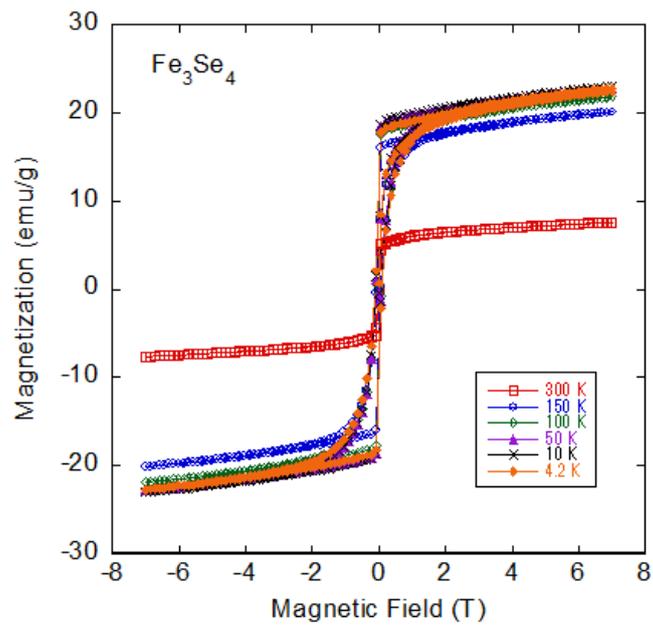

Fig. S3. Magnetization loops at 4.2-300 K for monoclinic $Fe_3Se_4$.